\documentclass{article}
\usepackage{setspace}

\begin{document}

\title{Self Force of a Charge in a Real Current Environment}
\date{}
\maketitle

Reuven Ianconescu

26 Rothenstreich Str., Tel-Aviv, Israel

r\_iancon@excite.com\\

L .P. Horwitz

School of Physics and Astronomy, Raymond and Beverly Sackler,

Faculty of Exact Sciences, Tel Aviv University,

Ramat Aviv 69978, Israel

larry@post.tau.ac.il\\

\noindent The analysis of the EM radiation from a single charge shows that the radiated power depends on the retarded acceleration of the charge. Therefore for consistency, an accelerated charge, free from the influence of external forces, should gradually lose its acceleration, until its total energy is radiated. Calculations show that the self force of a charge, which compensates for its radiation, is proportional to the derivative of the acceleration. However, when using this self force in the equation of motion of the charge, one gets a diverging solution, for which the acceleration runs away to infinity. This means that there is an inconsistency in the solution of the single charge problem. However, in the construction of the conserved Maxwell charge density, there is implicitly an integral over the corresponding world line which corresponds to a collection of charged spacetime events. One may therefore consistently think of the "self force'' as the force on a charge due to another charge at the retarded position. From this point of view, the energy is evidently conserved and the radiation process appears as an absorbing resistance to the feeding source. The purpose of this work is to learn about the behavior of single charges from the behavior of a real current, corresponding to the set of charges moving on a world line, and to study the analog of the self force of a charge associated with the radiation resistance of a continuum of charges.\\



\noindent{\bf\large 1. Introduction}\\

\noindent Antennas and EM radiation problems have been extensively discussed in the literature, and usually have closed form solutions, or at least closed numerically formulated solutions [1,2]. This means that the behavior of charges which are part of a real given electrical current is well described.

However, the behavior of single charges has still many open questions [3,4,5]. In the construction of the conserved Maxwell charge density, there is implicitly an integral [6] over the corresponding world line which corresponds to a collection of charged spacetime events. One may therefore consistently think of the "self force'' as the force on a charge due to another charge at the retarded position. In this paper we wish to deduce and extrapolate the behavior of single charges, from the behavior of a real electrical current. Specifically, we would like to understand if the "self" force of a charge, which acts as a radiation damping force, is an analogy to the radiation resistance, which acts like a power absorber from the feeding source.

We will use in this paper $\epsilon _{0}=1/(4\pi );\mu _{0}=4\pi ;c=1$, and put back the standard values only where it will be specifically mentioned.\\

\noindent{\bf\large 2. Formulation}\\

\noindent We formulate here the following problem:

\begin{picture}(170,170)(-50,0)
\linethickness{0.5 pt}
\put(125,0){\vector(0,1){150}}
\linethickness{2 pt}
\put(125,75){\line(0,1){65}}
\put(125,5){\line(0,1){65}}
\linethickness{0.5 pt}
\put(123,72.5){\line(1,0){4}}
\put(95,75){\line(1,0){31}}
\put(95,70){\line(1,0){31}}

\put(95,75){\vector(1,0){25}}
\put(126,70){\vector(-1,0){11}}

\put(85,72.5){\circle{18}}
\put(83,72.5){\oval(5,5)[t]}
\put(88,72.5){\oval(5,5)[b]}
\put(80,90){\makebox(20,5){$I_0\cos(\omega t)$}}
\put(130,150){\makebox(5,5){$z$}}
\put(132,135){\makebox(18,5){$L/2$}}
\put(132,5){\makebox(18,5){$-L/2$}}
\end{picture}


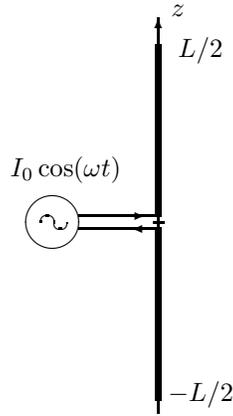
\begin{center} Figure 1: Short dipole antenna  \end{center}

The antenna is fed by the current source $ I(t) $ (at $ z=0 $), which can be expanded in a Fourier series of functions. Therefore, without loss of generality we may consider the harmonic excitation: $ I(t)=I_0\cos(\omega t) $. At an arbitrary $z$, the current will be:\\

$ I(z,t)=I_f(z,t)+I_b(z,t) $ \hfill (1) \\

\noindent where $I_f$, $I_b$ are the current waves which move forwards and backwards, respectively, in the $z$ direction, and have the form of:\\

$I_{f,b}(z,t)=C_{1,2}\cos\omega(t\mp z +\phi_{1,2})$ \hfill (2)\\

\noindent where the constants $C_{1,2}$, $\phi_{1,2}$ must satisfy the boundary conditions of zero current at $z= \pm L/2$ and $I_0 cos(\omega t)$ current at $z=0$. Those boundary conditions establish the spatial dependency of the current $f(z)$ so that the current in (1) may also be written as:\\

$I(z,t)=I_0\cos(\omega t)f(z)$, $f(0)=1$, $f(\pm L/2)=0$ \hfill (1a)\\

\noindent However, these boundary conditions are not important for the work of this paper because we are interested in locally analyzing the behavior of charges. But we will use the general form of the current, for establishing connections between the current and microscopic parameters, like speed and acceleration of the single charges. So we rewrite the current in (1) as:\\

$I(z,t)=C_1 \cos\omega(t-z +\phi_1) + C_2 \cos\omega(t+z +\phi_2)$ \hfill (3)\\

\noindent and\\

$\omega = 2\pi f= 2\pi/\lambda$ \hfill (4)\\

\noindent where $f$ is the frequency of the excitation and $\omega$ is the wavelength.

Next we have to calculate the speed of the charges $v(z,t)$.
Having a one dimensional configuration, it will be convenient to define $\rho$, the charge density, as the charge per unit of length. Therefore, to be consistent we have to define $J$, the current density, as the current $I$ itself:\\

$J(z,t)=I_f(z,t)+I_b(z,t)=C_1\cos\omega(t-z +\phi_1) + C_2\cos\omega(t+z +\phi_2)$ \hfill (5)\\

\noindent Knowing that:\\

$\partial_\mu J^\mu = 0$ therefore $\frac{\partial\rho}{\partial t} + \frac{\partial J}{\partial z}=0$ \hfill (6)\\

\noindent we calculate:\\

$\frac{\partial\rho}{\partial t} = \omega [-C_1\sin\omega(t-z +\phi_1) + C_2\sin\omega(t+z +\phi_2) ]$ \hfill (7)\\

\noindent Therefore:

$\rho = \omega \int dt [-C_1\sin\omega(t-z +\phi_1) + C_2\sin\omega(t+z +\phi_2) ]=$

$I_f(z,t) - I_b(z,t) + \rho_0$ \hfill (8)\\

\noindent $\rho_0$ is the integration constant and represents the density of the free electrons per unit of length in the conductor. As we shall see, $\rho_0\gg|I_f(z,t) - I_b(z,t)|$ therefore $I_f(z,t)$, $I_b(z,t)$ constitute low amplitude waves moving forward and backwards, respectively, around the average charge density $\rho_0$.

Knowing the relation $J=\rho v$, we obtain for $v(z,t)$:\\

$v(z,t)= \frac {J}{\rho} = \frac{I}{\rho} = \frac{I_f(z,t) + I_b(z,t)}{I_f(z,t) - I_b(z,t) + \rho_0}$ \hfill (9)\\

\noindent For any practical purpose $\rho_0$ is the dominant term in the denominator of (9), since $\rho_0 \gg C_{1,2}/c$ (we insert here the light velocity $c$). For example, if $C_{1,2} = 10A$, then $C_{1,2}/c = 1.87\times 10^{11}$ electrons/$m$. Let's say our wire has a surface of $0.5 mm^2$, so $C_{1,2}/c$ corresponds to $7.5\times 10^{17}$ electrons/$m^3$, while the charge density in copper is $8\times 10^{28}$ electrons/$m^3$. So $\rho_0$ is bigger than $C_{1,2}/c$ by 11 orders of magnitude, and the velocity of the electrons is smaller than $c$ by 11 orders of magnitude, i.e., about $10^{-4} m/sec$. We therefore may approximate the velocity of the charges by:\\

$v(z,t)= \frac{1}{\rho_0} I(z,t)$ \hfill (10)\\

\noindent{\bf\large 3. Discretization of the continuum}\\

\noindent The next step is to discretize the current into charges of magnitude, having a distanceone from the other, so that, as in Figure 2.\\

\begin{picture}(170,170)(-50,0)
\linethickness{0.5 pt}
\put(100,135){\vector(0,1){15}}
\put(105,150){\makebox(8,5){$z$}}

\linethickness{2 pt}
\put(80,0){\line(0,1){135}}
\put(120,0){\line(0,1){135}}
\put(100,90){\circle*{8}}
\put(100,65){\circle*{8}}
\put(100,40){\circle*{8}}
\put(100,15){\circle*{8}}

\linethickness{0.5 pt}
\put(40,140){\line(1,-1){15}}
\put(55,125){\line(1,1){10}}
\put(65,135){\vector(1,-1){13}}

\put(140,140){\line(-1,-1){35}}
\put(105,105){\line(2,-1){10}}
\put(115,100){\vector(-4,-3){10}}

\put(40,65){\vector(1,0){55}}
\put(40,40){\vector(1,0){55}}

\put(40,70){\makebox(25,5){Charge A}}
\put(40,45){\makebox(25,5){Charge B}}

\put(125,65){\line(1,0){25}}
\put(125,40){\line(1,0){25}}

\put(132,60){\vector(0,1){5}}
\put(132,45){\vector(0,-1){5}}

\put(134,50){\makebox(30,5){$\Delta z=q/\rho_0$}}

\put(40,145){\makebox(25,5){conductor}}
\put(143,145){\makebox(25,5){free charges}}
\put(147,137){\makebox(25,5){of size $q$}}

\end{picture}

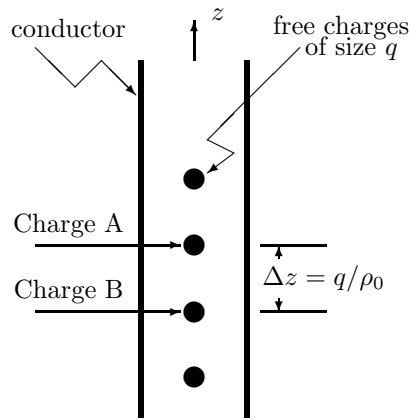
\begin{center} Figure 2: Close-up view of the conductor\end{center}



We will examine the forces acting on the "test charge" B (Fig. 2), as a result of a disturbance produced on charge A.

Before the disturbance occurred, all charges had a constant velocity, and on each charge, two opposite static forces of magnitude $q^2/(\Delta z)^2$ acted.

A disturbance which occurred at charge A at an earlier time $t=-\Delta t=-\Delta z$, appears as a disturbance perceived at charge B at $t=0$.

The motion of the "disturbed" charge A, $z_A(t)$ can be expanded for $t=-\Delta t$, up to third order, as follows:\\

$z_A(-\Delta t)=z_{A_0}-v\Delta t+\frac{1}{2} a \Delta t^2 - \frac{1}{6} \dot{a} \Delta t^3$ \hfill (11)\\

\noindent where $v$, $a$ and $\dot{a}$ are the velocity, acceleration and its derivative, measured at $t=0$. The velocity of the "disturbed" charge A is:\\

$v_A(-\Delta t)=v- a \Delta t + \frac{1}{2} \dot{a} \Delta t^2$ \hfill (12)\\

\noindent The full distance to $z_B$ from the disturbed charge at A must arrive at light velocity, therefore, $\Delta z$ at time $-\Delta t$ is:\\

$\Delta z(-\Delta t)=z_B-z_A(-\Delta t)=\Delta z_0+v\Delta t-\frac{1}{2} a \Delta t^2 + \frac{1}{6} \dot{a} \Delta t^3=\Delta t$ \hfill (13)\\

\noindent Formula (13) tells us that the retarded distance equals the time needed to propagate the disturbance, and expands this distance in powers of $\Delta t$, where $\Delta z_0$ is the distance between the charges A and B at $t=0$.

The near field of a charge is expressed by [4]:\\

$E_z = \frac{q}{[R_\mu v^\mu]^2}_{\tt RET}=\frac{q}{[\gamma\Delta z(1-v_A)]^2}_{|_{-\Delta t}}$, \hfill (14)\\

\noindent where $\Gamma$ is very close to $1$. We will use (12) and (13) for putting the values of $\Delta z$ and $v_A$ into (14) and get from (13):\\

$\Delta z_0 - \Delta t (1 - v + \frac{1}{2} a \Delta t - \frac{1}{6} \dot{a} \Delta t^2)=0$ \hfill (15)\\

\noindent With the aid of (15), we obtain from (12)\\

$1-v_A(-\Delta t)=\frac{\Delta z_0}{\Delta t} - \frac{1}{2} a \Delta t + \frac{1}{6} \dot{a} \Delta t^2 + a \Delta t - \frac{1}{2} \dot{a} \Delta t^2 \simeq 1 + \frac{1}{2} a \Delta t - \frac{1}{3} \dot{a} \Delta t^2$ \hfill (16)\\

\noindent Here, we have used the approximation $\Delta z_0 \simeq \Delta t$ because the velocity is very small.

Using this result, $\Delta z=\Delta t$ and $\gamma\simeq 1$ in (14), we get the field at charge B, caused by a disturbance which occurred $\Delta t$ earlier at charge A, as\\

$E_z = q (\frac{1}{\Delta t^2}-  \frac{a}{\Delta t} + \frac{2}{3} \dot{a} - \frac{1}{4} a^2)$ \hfill (17)\\

\noindent The first term $q/\Delta t^2$ {\it is always cancelled} by the force of the "other" neighbor, as mentioned before, it therefore {\it can be ignored}.

The third term is identical to what is considered to be the field which creates the self-force of a charge, but here it was derived as the force on a charge, due to a disturbance on a neighboring charge. As we shall see, it is the {\it only} term which creates the radiation resistance (which we see is a local phenomenon on the world line).

We will call the last 3 terms of (17) $E_{damp}$, and after replacing $\Delta t$ by $\Delta z$, we obtain:\\

$E_{damp} = q (-{\frac{a}{\Delta z}} + \frac{2}{3} \dot{a} - \frac{1}{4} a^2)$ \hfill (18)\\

\noindent The potential difference $V$ on a wire segment of length $\Delta z$, resulting from $E_{damp}$, will be denoted as the "damping" tension (or "damping" voltage) and is calculated by:\\

$V=-E_{damp}\Delta z$ \hfill (19)\\

\noindent The magnitude $\Delta z$ is according to the discretization defined above, so that\\

$\Delta z=q/\rho_0$ \hfill (20)\\

\noindent We therefore get from (18), (19) and (20):

$V = -{\frac{q^2 [2/3 \ddot{v}-1/4 (\dot{v})^2]}{\rho_0}}  + q \dot{v}$ \hfill (21)\\

\noindent According to (10):\\

$\dot{v}=\frac{1}{\rho_0} \frac{\partial I}{\partial t}$ \ and \ $\ddot{v}=\frac{1}{\rho_0}\frac{\partial^2 I}{\partial t^2}$ \hfill (22)\\

\noindent Dealing with harmonic excitation, $\partial / \partial t$ is like multiplication by $\omega$ (up to a $90^0$ phase). So $\ddot{v} \sim \omega^2 v$, and $(\dot{v})^2 \sim (\omega v)^2$. Having $v$ smaller by 11 orders of magnitude than light velocity, $\dot{v}^2$ is completely negligible relative to $\ddot{v}$.

We therefore may write (21) (after neglecting the $\dot{v}^2$ term), using (22):\\

$V =- \frac{2}{3} (\frac{q}{\rho_0})^2 \frac{\partial^2 I}{\partial t^2} + \frac{q}{\rho_0} \frac{\partial I}{\partial t} = - \frac{2}{3} \Delta z^2 \frac{\partial^2 I}{\partial t^2} + \Delta z \frac{\partial I}{\partial t}$ \hfill (23)\\

\noindent The power radiated by the segment $\Delta z$ is $\Delta P=VI$ and given by:\\

$\Delta P= -\frac{2}{3} \frac{\partial^2 I}{\partial t^2} I \Delta z^2 + \frac{\partial I}{\partial t} I \Delta z$ \hfill (24)\\

\noindent According to (3), (and also because $I$ satisfies the wave equation):\\

$\frac{\partial^2 I}{\partial t^2} = -\omega^2 I=-(\frac{2\pi}\lambda)^2 I$ \hfill (25)\\

\noindent The constant ratio between $\partial^2 I / \partial t^2$ and $I$ means that the current is "in phase" with its second derivative, and therefore the first part of $\Delta P$ in (24), integrated over time, represents radiated energy. However, the multiplication of $I$ by $\partial I / \partial t$ has the form of $\cos\omega(t-z +\phi)\sin\omega(t-z +\phi)$ and therefore the second part of $\Delta P$ represents a {\it reactive} power, which results in zero energy after integrating on an integer number of time cycles. The reactive power represents power which is returned to the source each time cycle. We are therefore interested in the first term of $\Delta P$ in (24).

Putting (25) into the first part of (24) we obtain:\\

$\Delta P(z)= \frac{2}{3} (\frac{2\pi}{\lambda})^2 \frac{1}{2} [I(z)\Delta z]^2$ \hfill (26)\\

\noindent In (26) we omitted the time dependence, which is harmonic (see (1a)), and multiplied by $\frac{1}{2}$ which is $(1/T)\int^{T}_{0} \cos^2 (2 \pi/T)t dt$. So (26) represents the mean emitted power over a time cycle $T=2 \pi/\omega$.

Therefore the mean power emitted from the whole wire is:\\

$P= \frac{2}{3} (\frac{2\pi}{\lambda})^2 \frac{1}{2} [\int^{L/2}_{-L/2} I(z,t) dz]^2 \equiv \frac{2}{3} (\frac{2\pi}{\lambda})^2 \frac{1}{2} [I_{av} L]^2=\frac{2}{3} ({2\pi})^2 (\frac{L}{\lambda})^2 \frac{1}{2} I^2_{av}$ \hfill (27)\\

\noindent Here we defined $I_{av} \equiv (1/L)\int^{L/2}_{-L/2} I(z) dz$.

The radiation resistance is defined by:\\

$R \equiv \frac{2P}{I_0^2} = \frac{2}{3} ({2\pi})^2 (\frac{L}{\lambda})^2 (I_{av} / I_0)^2$. \hfill (28)\\

The factor $2$ is due to $P$ being averaged on a time cycle, and $I_0$ is the amplitude. Now we put back the units: $Z_0 = 1/(\epsilon_0 c) = \sqrt {\mu_0 / \epsilon_0} = 120 \pi$ would be the impedance of the free space, if we had used $\epsilon_0=1$; but we used $\epsilon_0=1/4\pi$, so we have to multiply (28) by $Z_0 = \frac{120\pi}{4\pi}=30$. We therefore obtain the radiation resistance:\\

$R=80\pi^2 (\frac{L}{\lambda})^2 (I_{av} / I_0)^2 \Omega$ \hfill (29)\\

\noindent which is identical to the result in the literature [1,2].\\

\noindent{\bf\large 4. Discussion}\\

The radiation resistance  of two particular cases can be immediately derived from (29).

One is a halfwave dipole $L=\lambda /2$, for which the boundary conditions give $f(z)=\cos(2\pi z/\lambda)$ in (1a). Here we obtain $I_{av}=(2/\pi)I_0$, hence $R=80\Omega$.

The second is an ideal case of very short dipole antenna, in which we suppose $I_{av}=I_0$, (as if the boundary conditions were satisfied by two discs at the ends of the wire to act as charge reservoirs). In this case we obtain $R=80\pi^2 (L/\lambda)^2\Omega$. This result could have been obtained by directly dividing the voltage in the first part of (23) by the current $I\equiv I_0$, and obtain $R=80\pi^2(\Delta z/\lambda)^2\Omega$, if $L=\Delta z$.

This is actually the most interesting case for our study, because it considers the segment $\Delta z$ (which contains the charge $q$) as an antenna. This segment has the radiation resistance $R=(2/3)(2\pi)^2(\Delta z/\lambda)^2$ (in our units), which directly results from the third term of (17), and could also be written as:\\

$R=\frac{P_{self}}{I^2}$, \hfill (30)\\

\noindent where the work is $P_{self}=F_{self} v$, $F_{self}$ being $(2/3)q^2\dot{a}$, and $I$ is the charge flow per unit time - or current. We see that $F_{self}$ in a continuum of charges could be considered as the force acted by the perturbed neighbor.

The radiation resistance is therefore the "self work" divided by the charges flow rate squared.

We have to remark that this result has been derived with the aid of several approximations like $\rho_0\gg C_{1,2}/c$, resulting in $v\ll c$, which are {\it very} accurate for charges in a real wire, but cannot be counted on for a single charge. Therefore, result (30) is not a proof of the formula $F_{self}=(2/3)q^2\dot{a}$, but rather shows that in a {\it macro} perspective, it {\it looks like} the self force of a charge, which accounts for the radiation, is $(2/3)q^2\dot{a}$.

Moreover, $P_{self}$ comes out to be also $(2/3)q^2 a^2$ (averaged on a time cycle), which is the Larmor formula, describing the emitted energy per unit of emission time. This is because: $v(t)=\frac{I(t)}{\rho_0}=\frac{I_0\cos(\omega t)}{\rho_0}\equiv v_0\cos(\omega t)$, $a(t)=-\omega v_0\sin(\omega t)$, $\dot{a}(t)= -\omega^2 v_0\cos(\omega t)$ and therefore $\dot{a}v= -(\omega v_0)^2\cos^2(\omega t)$ and $a^2=(\omega v_0)^2\sin^2(\omega t)$. Both $\sin^2(\omega t)$ and $\cos^2(\omega t)$ averaged on a cycle, give $1/2$.

The result that $|\dot{a}v|=a^2$ implies that $P_{self}$ represents the "self work", i.e. $F_{self}v$ and also the emitted energy per unit of emission time, suggesting that the self force is an energy conservation force.\\

\noindent{\bf\large References}\\

\begin{enumerate}
\item Kai Fong Lee, {\it Principles of Antenna Theory}, John Wiley \& Sons Ltd, 1984.
\item G. Dubost, {\it Flat Radiating Dipoles and Applications to Arrays}, John Wiley \& Sons Ltd, 1981.
\item P.A.M. Dirac, {\it Classical theory of radiating electrons}, Proc. R. Soc. London Ser. A 167 (1938) 148.
\item F. Rohrlich, {\it Classical Charged Particles}, Addison-Wesley, 1965.
\item R. Ianconescu and L.P. Horwitz, {\it Self-force of a classical charged particle}, Physical Review A, Vol. 45, Number 7, pp. 4346, 1992.
\item J.D. Jackson, {\it Classical Electrodynamics}, 2nd ed., Wiley, 1975.
\end{enumerate}
\end{document}